\documentclass[runningheads]{llncs}
\usepackage{graphicx}
\usepackage{amsfonts}
\usepackage{makecell}
\usepackage[marginal]{footmisc}

\begin{document}
\title{Swin-Unet: Unet-like Pure Transformer for Medical Image Segmentation
}
%
%
\author{Hu Cao\inst{1\dagger} \and 
Yueyue Wang\inst{2\dagger} \and
Joy Chen\inst{1} \and
Dongsheng Jiang\inst{3*}
\and
Xiaopeng Zhang\inst{3*}
\and
Qi Tian\inst{3*}
\and
Manning Wang\inst{2}
}

\authorrunning{Hu Cao et al.}
%

\institute{Technische Universit\"at M\"unchen, M\"unchen, Germany \and
Fudan University, Shanghai, China \and
Huawei Technologies, Shanghai, China 
}

\maketitle              

\begin{abstract}
In the past few years, convolutional neural networks (CNNs) have achieved milestones in medical image analysis. Especially, the deep neural networks based on U-shaped architecture and skip-connections have been widely applied in a variety of medical image tasks. However, although CNN has achieved excellent performance, it cannot learn global and long-range semantic information interaction well due to the locality of convolution operation. In this paper, we propose Swin-Unet, which is a Unet-like pure Transformer for medical image segmentation. The tokenized image patches are fed into the Transformer-based U-shaped Encoder-Decoder architecture with skip-connections for local-global semantic feature learning. Specifically, we use hierarchical Swin Transformer with shifted windows as the encoder to extract context features. And a symmetric Swin Transformer-based decoder with patch expanding layer is designed to perform the up-sampling operation to restore the spatial resolution of the feature maps. Under the direct down-sampling and up-sampling of the inputs and outputs by $4\times$, experiments on multi-organ and cardiac segmentation tasks demonstrate that the pure Transformer-based U-shaped Encoder-Decoder network outperforms those methods with full-convolution or the combination of transformer and convolution. The codes and trained models will be publicly available at https://github.com/HuCaoFighting/Swin-Unet.
 
\end{abstract}
\footnote{*Corresponding author}
\footnote{$\dagger$ Work done as an intern in Huawei Technologies}

\section{Introduction}

Benefiting from the development of deep learning, computer vision technology has been widely used in 
medical image analysis. Image segmentation is an important part of medical image analysis. In particular, accurate and robust medical image segmentation can play a cornerstone role in computer-aided diagnosis and image-guided clinical surgery~\cite{unetr,trans-unet}. 

Existing medical image segmentation methods mainly rely on fully convolutional neural network (FCNN) with U-shaped structure~\cite{unet,nnu-net,RA-UNet}. The typical U-shaped network, U-Net~\cite{unet}, consists of a symmetric Encoder-Decoder with skip connections. In the encoder, a series of convolutional layers and continuous down-sampling layers are used to extract deep features with large receptive fields. Then, the decoder up-samples the extracted deep features  to the input resolution for pixel-level semantic prediction, and the high-resolution features of different scale from the encoder are fused with skip connections to alleviate the loss of spatial information caused by down-sampling. With such an elegant structural design, U-Net has achieved great success in a variety of medical imaging applications. Following this technical route, many algorithms such as 3D U-Net~\cite{3D-UNet}, Res-UNet~\cite{res-unet}, U-Net++~\cite{Unet++} and UNet3+~\cite{unet3+} have been developed for image and volumetric segmentation of various medical imaging modalities. The excellent performance of these FCNN-based methods in 
cardiac segmentation, organ segmentation and lesion segmentation proves that CNN has a strong ability of learning discriminating features.

Currently, although the CNN-based methods have achieved excellent performance in the field of medical image segmentation, they still cannot fully meet the strict requirements of medical applications for segmentation accuracy. Image segmentation is still a challenge task in medical image analysis. Since the intrinsic locality of convolution operation, it is difficult for CNN-based approaches to learn explicit global and long-range semantic information interaction~\cite{trans-unet}. Some studies have tried to address this problem by using atrous convolutional layers~\cite{deeplab,cenet}, self-attention mechanisms~\cite{att-gate,nonlocal}, and image pyramids~\cite{pyramid}. However, these methods still have limitations in modeling long - range dependencies. Recently, inspired by Transformer's great success in the nature language processing (NLP) domain~\cite{transformer}, researchers have tried to bring Transformer into the vision domain~\cite{detr}. In~\cite{vit}, vision transformer (ViT) is proposed to perform the image recognition task. Taking 2D image patches with positional embeddings as inputs and pre-training on large dataset, ViT achieved comparable performance with the CNN-based methods. Besides, data-efficient image transformer (DeiT) is presented in~\cite{deit}, which indicates that Transformer can be trained on mid-size datasets and that a more robust Transformer can be obtained by combining it with the distillation method. In~\cite{swin}, a hierarchical Swin Transformer is developed. Take Swin Transformer as vision backbone, the authors of~\cite{swin} achieved state-of-the-art performance on Image classification, object detection and semantic segmentation. The success of ViT, DeiT and Swin Transformer in image recognition task demonstrates the potential for Transformer to be applied in the vision domain.

Motivated by the Swin Transformer's~\cite{swin} success, we propose Swin-Unet to leverage the power of Transformer for 2D medical image segmentation in this work. To our best knowledge, Swin-Unet is a first pure Transformer-based U-shaped architecture that consists of encoder, bottleneck, decoder, and skip connections. Encoder, bottleneck and decoder are all built based on Swin Transformer block~\cite{swin}. The input medical images are split into non-overlapping image patches. Each patch is treated as a token and fed into the Transformer-based encoder to learn deep feature representations. The extracted context features are then up-sampled by the decoder with patch expanding layer, and fused with the multi-scale features from the encoder via skip connections, so as to restore the spatial resolution of the feature maps and further perform segmentation prediction.  Extensive experiments on multi-organ and cardiac segmentation datasets indicate that the proposed method has excellent segmentation accuracy and robust generalization ability. Concretely, our contributions can be summarized as: (1) Based on Swin Transformer block, we build a symmetric Encoder-Decoder architecture with skip connections. In the encoder, self-attention from local to global is realized; in the decoder, the global features are up-sampled to the input resolution for corresponding pixel-level segmentation prediction. (2) A patch expanding layer is developed to achieve up-sampling and feature dimension increase without using convolution or interpolation operation. (3) It is found in the experiment that skip connection is also effective for Transformer, so a pure Transformer-based U-shaped Encoder-Decoder architecture with skip connection is finally constructed, named Swin-Unet.

\section{Related work}

\subsubsection{CNN-based methods}:
Early medical image segmentation methods are mainly contour-based and traditional machine learning-based algorithms~\cite{shape,markov}. With the development of deep CNN, U-Net is proposed in~\cite{unet} for medical image segmentation. Due to the simplicity and superior performance of the U-shaped structure, various Unet-like methods are constantly emerging, such as Res-UNet~\cite{res-unet}, Dense-UNet~\cite{denseunet}, U-Net++~\cite{Unet++} and UNet3+~\cite{unet3+}. And it is also introduced into the field of 3D medical image segmentation, such as 3D-Unet~\cite{3D-UNet} and V-Net~\cite{VNet}. At present, CNN-based methods have achieved tremendous success in the field of medical image segmentation due to its powerful representation ability.

\subsubsection{Vision transformers}:
Transformer was first proposed for the machine translation task in~\cite{transformer}. In the NLP domain, the Transformer-based methods have achieved the state-of-the-art performance in various tasks~\cite{bert}. Driven by Transformer's success, the researchers introduced a pioneering vision transformer (ViT) in~\cite{vit}, which achieved the impressive speed-accuracy trade-off on image recognition task.
Compared with CNN-based methods, the drawback of ViT is that it requires pre-training on its own large dataset. To alleviate the difficulty in training ViT, Deit~\cite{deit} describes several training strategies that allow ViT to train well on ImageNet. Recently, several excellent works have been done baed on ViT~\cite{pvt,tit,swin}. It is worth mentioning that an efficient and effective hierarchical vision Transformer, called Swin Transformer, is proposed as a vision backbone in~\cite{swin}. Based on the shifted windows mechanism, Swin Transformer achieved the state-of-the-art performance on various vision tasks including image classification, object detection and semantic segmentation. In this work, we attempt to use Swin Transformer block as basic unit to build a U-shaped Encoder-Decoder architecture with skip connections for medical image segmentation, thus providing a benchmark comparison for the development of Transformer in the medical image field.

\subsubsection{Self-attention/Transformer to complement CNNs}:
In recent years, researchers have tried to introduce self-attention mechanism into CNN to improve the performance of the network~\cite{nonlocal}. In~\cite{att-gate}, the skip connections with additive attention gate are integrated in U-shaped architecture to perform medical image segmentation. However, this is still the CNN-based method. Currently, some efforts are being made to combine CNN and Transformer to break the dominance of CNNs in medical image segmentation~\cite{trans-unet,medical-transformer,unetr}. In~\cite{trans-unet}, the authors combined Transformer with CNN to constitute a strong encoder for 2D medical image segmentation. Similar to~\cite{trans-unet},~\cite{medical-transformer} and~\cite{transfuse}  use the complementarity of Transformer and CNN to improve the segmentation capability of the model. Currently, various combinations of Transformer with CNN are applied in multi-modal brain tumor segmentation~\cite{transbts} and 3D medical image segmentation~\cite{unetr,cotr}. Different from the above methods, we try to explore the application potential of pure Transformer in medical image segmentation.

\section{Method}

\subsection{Architecture overview}

\begin{figure}[t!]
\includegraphics[width=\textwidth]{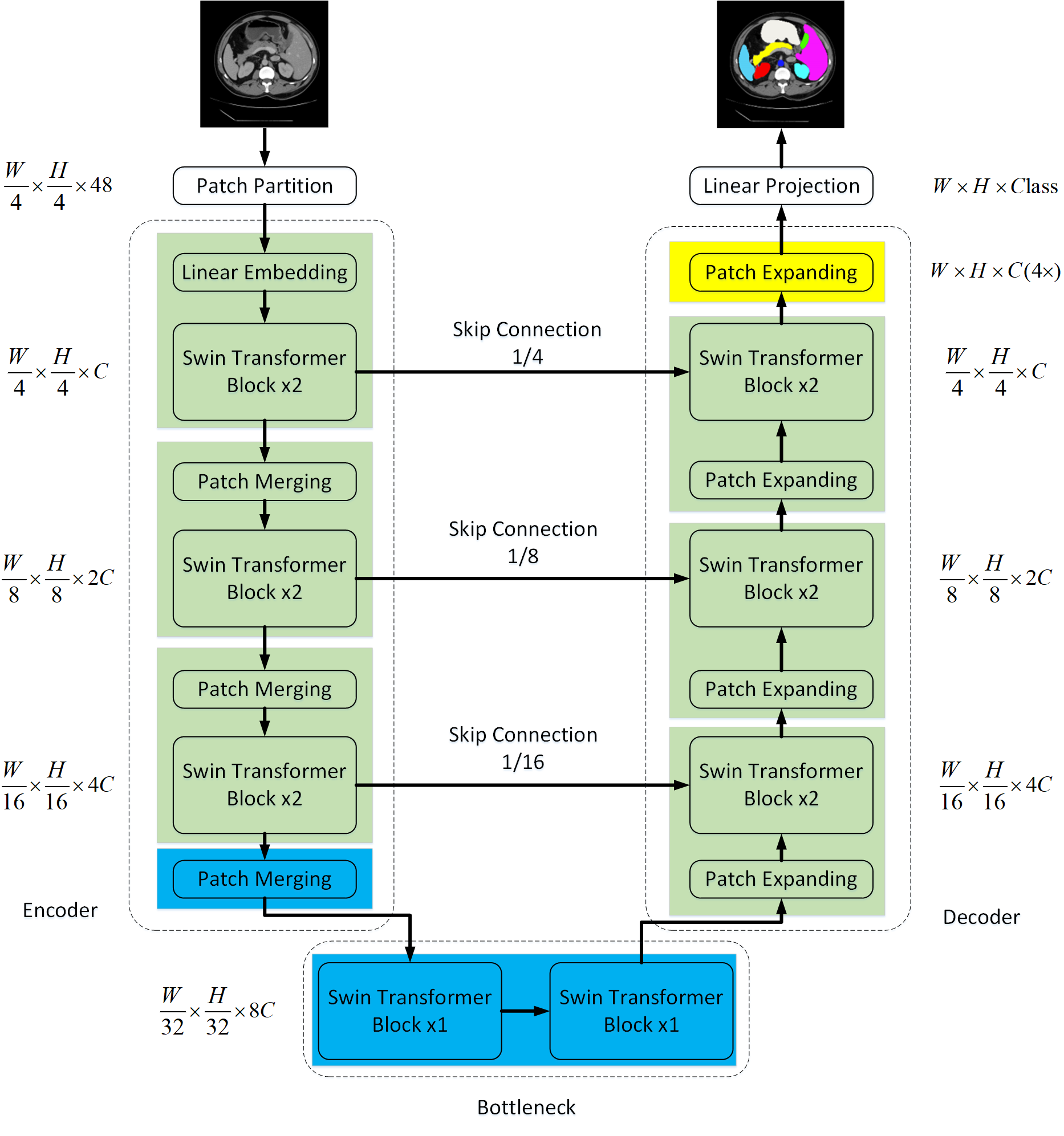}
\caption{The architecture of Swin-Unet, which is composed of encoder, bottleneck, decoder and skip connections. Encoder, bottleneck and decoder are all constructed based on swin transformer block.} \label{swin_unet}
\end{figure}

The overall architecture of the proposed Swin-Unet is presented in Figure.~\ref{swin_unet}. Swin-Unet consists of encoder, bottleneck, decoder and skip connections. The basic unit of Swin-Unet is Swin Transformer block~\cite{swin}. For the encoder, to transform the inputs into sequence embeddings, the medical images are split into non-overlapping patches with patch size of $4\times4$. By such partition approach, the feature dimension of each patch becomes to $4\times4\times3=48$. Furthermore, a linear embedding layer is applied to projected feature dimension into arbitrary dimension (represented as C). The transformed patch tokens pass through several Swin Transformer blocks and patch merging layers to generate the hierarchical feature representations. Specifically, patch merging layer is responsible for down-sampling and increasing dimension, and Swin Transformer block is responsible for feature representation learning.
Inspired by U-Net~\cite{unet}, we design a symmetric transformer-based decoder. The decoder is composed of Swin Transformer block and patch expanding layer. The extracted context features are fused with multiscale features from encoder via skip connections to complement the loss of spatial information caused by down-sampling. In contrast to patch merging layer, a patch expanding layer is specially designed to perform up-sampling. The patch expanding layer reshapes feature maps of adjacent dimensions into a large feature maps with $2\times$ up-sampling of resolution. In the end, the last patch expanding layer is used to perform $4\times$ up-sampling to restore the resolution of the feature maps to the input resolution ($W\times H$), and then a linear projection layer is applied on these up-sampled features to output the pixel-level segmentation predictions. We would elaborate each block in the following

\subsection{Swin Transformer block}
\begin{figure}[t!]
\centering
\includegraphics[width=0.8\textwidth]{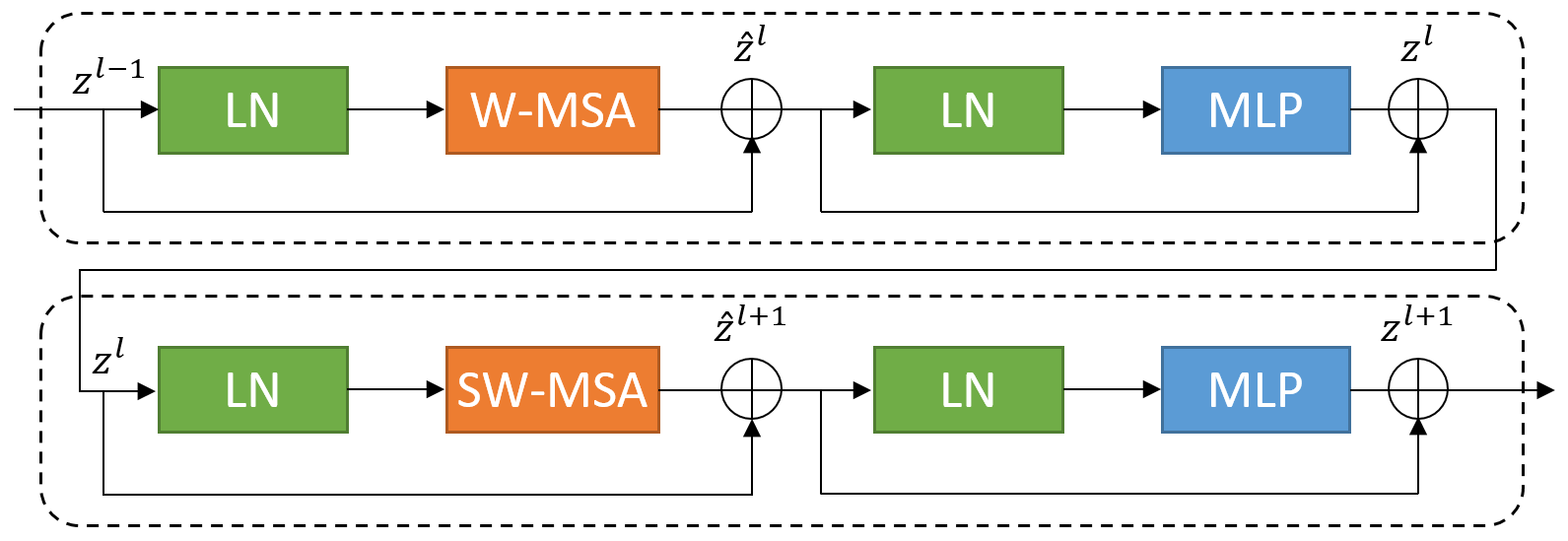}
\caption{Swin transformer block.} \label{swin_block}
\end{figure}

Different from the conventional multi-head self attention (MSA) module, swin transformer block~\cite{swin} is constructed based on shifted windows. In Figure.~\ref{swin_block}, two consecutive swin transformer blocks are presented. Each swin transformer block is composed of LayerNorm (LN) layer, multi-head self attention module, residual connection and 2-layer MLP with GELU non-linearity. The window-based multi-head self attention (W-MSA) module and the shifted window-based multi-head self attention (SW-MSA) module are applied in the two successive transformer blocks, respectively. Based on such window partitioning mechanism, continuous swin transformer blocks can be formulated as:

\begin{equation}
\hat{z}^l = {W\raisebox{0mm}{-}MSA}(LN(z^{l-1})) + z^{l-1}, 
\end{equation}

\begin{equation}
z^l = MLP(LN(\hat{z}^l))  + \hat{z}^{l}, 
\end{equation}

\begin{equation}
\hat{z}^{l+1} = {SW\raisebox{0mm}{-}MSA}(LN(z^l))  + z^l,  
\end{equation}

\begin{equation}
z^{l+1} = MLP(LN(\hat{z}^{l+1})) + \hat{z}^{l+1}, 
\end{equation}
where $\hat{z}^l$ and $z^l$ represent the outputs of the (S)W-MSA module and the MLP module of the $l^{th}$ block, respectively. Similar to the previous works~\cite{relation,local}, self-attention is computed as follows:

\begin{equation}
Attention(Q,K,V) = SoftMax(\frac{QK^T}{\sqrt{d}} + B)V , 
\end{equation}
where $Q,K,V\in \mathbb{R}^{M^2\times d}$ denote the query, key and value matrices. $M^2$ and $d$ represent the number of patches in a window and the dimension of the $query$ or $key$, respectively. And,  the values in $B$ are taken from the bias matrix $\hat{B}\in \mathbb{R}^{(2M-1)\times(2M+1)}$. 

\subsection{Encoder}

In the encoder, 
the C-dimensional tokenized inputs with the resolution of $\frac{H}{4}\times\frac{W}{4}$ are fed into the two consecutive Swin Transformer blocks to perform representation learning, in which the feature dimension and resolution remain unchanged. Meanwhile, 
the patch merging layer will reduce the number of tokens ($2\times$ down-sampling) and increase the feature dimension to $2\times$ the original dimension. This procedure will be repeated three times in the encoder.

\subsubsection{Patch merging layer}:
The input patches are divided into 4 parts and concatenated together by the patch merging layer. With such processing, the feature resolution will be down-sampled by $2\times$. And, since the concatenate operation results the feature dimension increasing by $4\times$, a linear layer is applied on the concatenated features to unify the feature dimension to the $2\times$ the original dimension.

\subsection{Bottleneck}
Since Transformer is too deep to be converged~\cite{deeper}, only two successive Swin Transformer blocks are used to constructed the bottleneck to learn the deep feature representation. In the bottleneck, the feature dimension and resolution are kept unchanged.

\subsection{Decoder}

Corresponding to the encoder, the symmetric decoder is built based on Swin Transformer block. To this end, in contrast to the patch merging layer used in the encoder, we use the patch expanding layer in the decoder to up-sample the extracted deep features. The patch expanding layer reshapes the feature maps of adjacent dimensions into a higher resolution feature map ($2\times$ up-sampling) and reduces the feature dimension to half of the original dimension accordingly. 

\subsubsection{Patch expanding layer}:
Take the first patch expanding layer as an example, before up-sampling, a linear layer is applied on the input features ($\frac{W}{32}\times \frac{H}{32} \times 8C$) to increase the feature dimension to $2\times$ the original dimension ($\frac{W}{32}\times \frac{H}{32} \times 16C$). Then, we use rearrange operation to expand the resolution of the input features to $2\times$ the input resolution and reduce the feature dimension to quarter of the input dimension ($\frac{W}{32}\times \frac{H}{32} \times 16C$ $\rightarrow$ $\frac{W}{16}\times \frac{H}{16} \times 4C$). We will discuss the impact of using patch expanding layer to perform up-sampling in section~\ref{upsampling}.

\subsection{Skip connection}
Similar to the U-Net~\cite{unet}, the skip connections are used to fuse the multi-scale features from the encoder with the up-sampled features. We concatenate the shallow features and the deep features together to reduce the loss of spatial information caused by down-sampling. Followed by a linear layer, the dimension of the concatenated features is remained the same as the dimension of the up-sampled features. In section~\ref{skip_num}, we will detailed discuss the impact of the number of skip connections on the performance of our model.
\section{Experiments}
\subsection{Datasets}
\textbf{Synapse multi-organ segmentation dataset (Synapse):} the dataset includes 30 cases with 3779 axial abdominal clinical CT images. Following~\cite{trans-unet,domain_adaptive}, 18 samples are divided into the training set and 12 samples into testing set. And the average Dice-Similarity coefficient (DSC) and average Hausdorff Distance (HD) are used as evaluation metric to evaluate our method on 8 abdominal organs (aorta, gallbladder, spleen, left kidney, right kidney, liver, pancreas, spleen, stomach). 


\textbf{Automated cardiac diagnosis challenge dataset (ACDC):} the ACDC dataset is collected from different patients using MRI scanners. For each patient MR image, left ventricle (LV), right ventricle (RV) and myocardium (MYO) are labeled. The dataset is split into 70 training samples, 10 validation samples and 20 testing samples. Similar to~\cite{trans-unet}, only average DSC is used to evaluate our method on this dataset.

\subsection{Implementation details}

The Swin-Unet is achieved based on Python 3.6 and Pytorch 1.7.0. For all training cases, data augmentations such as flips and rotations are used to increase data diversity. The input image size and patch size are set as $224\times224$ and $4$, respectively. We train our model on a Nvidia V100 GPU with 32GB memory. The weights pre-trained on ImageNet are used to initialize the model parameters. During the training period, the batch size is 24 and the popular SGD optimizer with momentum 0.9 and weight decay 1e-4 is used to optimize our model for back propagation. 

\begin{figure}[t!]
\includegraphics[width=\textwidth]{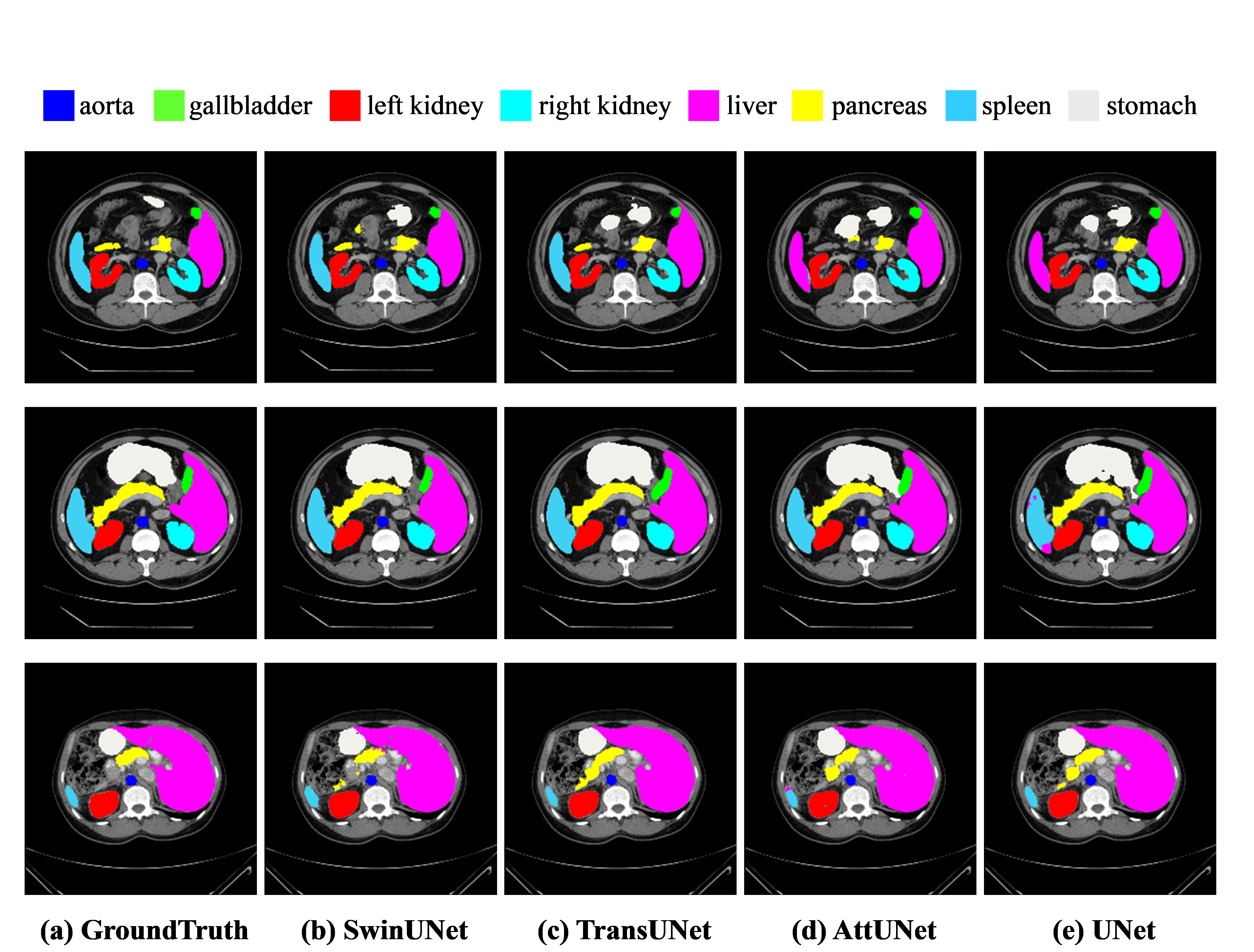}
\caption{The segmentation results of different methods on the Synapse multi-organ CT dataset.} \label{segmentation}
\end{figure}

\begin{table*}[t!]
\centering
\caption{Segmentation accuracy of different methods on the Synapse multi-organ CT dataset.}
\footnotesize
\resizebox{\textwidth}{!}{
\begin{tabular}{c|cc|cccccccc}
\hline
Methods &  DSC$\uparrow$ &HD$\downarrow$ & Aorta& Gallbladder& Kidney(L)& Kidney(R)& Liver& Pancreas& Spleen& Stomach\\
\hline
V-Net~\cite{V-Net} &  68.81 & -& 75.34&51.87&77.10&\textbf{80.75}&87.84&40.05&80.56&56.98\\
DARR~\cite{darr} &   69.77 & -& 74.74&53.77&72.31&73.24&94.08&54.18&89.90&45.96\\
R50 U-Net~\cite{trans-unet} & 74.68 & 36.87& 87.74&63.66&80.60&78.19&93.74&56.90&85.87&74.16 \\
U-Net~\cite{unet} & 76.85 & 39.70& 89.07&\textbf{69.72}&77.77&68.60&93.43&53.98&86.67&75.58 \\
R50 Att-UNet~\cite{trans-unet}  & 75.57&  36.97& 55.92&63.91&79.20&72.71&93.56&49.37&87.19&74.95 \\
Att-UNet~\cite{att-unet} & 77.77 & 36.02& \textbf{89.55}&68.88&77.98&71.11&93.57&\textbf{58.04}&87.30&75.75 \\
R50 ViT~\cite{trans-unet} & 71.29 &  32.87& 73.73&55.13&75.80&72.20&91.51&45.99&81.99&73.95\\
TransUnet~\cite{trans-unet} & 77.48 &  31.69&87.23&63.13&81.87&77.02&94.08&55.86&85.08&75.62 \\
\hline
SwinUnet & \textbf{79.13} & \textbf{21.55}&85.47&66.53&\textbf{83.28}&79.61&\textbf{94.29}&56.58&\textbf{90.66}&\textbf{76.60}\\
\hline
\end{tabular}
}
\label{synapse}
\end{table*}


\subsection{Experiment results on Synapse dataset}
The comparison of the proposed Swin-Unet with previous state-of-the-art methods on the Synapse multi-organ CT dataset is presented in Table.~\ref{synapse}. Different from TransUnet~\cite{trans-unet}, we add the test results of our own implementations of U-Net~\cite{unet}and Att-UNet~\cite{att-unet} on the Synapse dataset. Experimental results demonstrate that our Unet-like pure transformer method achieves the best performance with segmentation accuracy of $79.13\%$(DSC$\uparrow$) and $21.55\%$(HD$\downarrow$). Compared with Att-Unet~\cite{att-unet} and the recently method TransUnet~\cite{trans-unet}, although our algorithm did not improve much on the DSC evaluation metric, we achieved accuracy improvement of about $4\%$ and $10\%$ on the HD evaluation metric, which indicates that our approach can achieve better edge predictions.
The segmentation results of different methods on the Synapse multi-organ CT dataset are shown in Figure.~\ref{segmentation}. It can be seen from the figure that CNN-based methods tend to have over-segmentation problems, which may be caused by the locality of convolution operation. 
In this work, we demonstrate that by integrating Transformer with a U-shaped architecture with skip connections, the pure Transformer approach without convolution can better learn both global and long-range semantic information interactions, resulting in better segmentation results.

\begin{table}[t!]
\caption{Segmentation accuracy of different methods on the ACDC dataset.}\label{acdc}
\begin{tabular}{c|c|ccc}
\hline
Methods & DSC& RV & Myo& LV\\
\hline
R50 U-Net & 87.55 & 87.10&80.63& 94.92\\
R50 Att-UNet  & 86.75& 87.58&79.20&93.47 \\
R50 ViT & 87.57&86.07&81.88&94.75\\
TransUnet & 89.71&88.86&84.53&95.73 \\
\hline
SwinUnet & \textbf{90.00} &88.55&\textbf{85.62}&\textbf{95.83}\\
\hline
\end{tabular}
\centering
\end{table}

\subsection{Experiment results on ACDC dataset}

Similar to the Synapse dataset, the proposed Swin-Unet is trained on ACDC dataset to perform medical image segmentation. The experimental results are summarized in Table.~\ref{acdc}. By using the image data of MR mode as input, Swin-Unet is still able to achieve excellent performance with an accuracy of $90.00\%$, which shows that our method has good generalization ability and robustness.

\subsection{Ablation study}

In order to explore the influence of different factors on the model performance, we conducted ablation studies on Synapse dataset. Specifically, up-sampling, the number of skip connections, input sizes, and model scales are discussed below.

\begin{table}[t!]
\caption{Ablation study on the impact of the up-sampling}\label{up}
\footnotesize
\resizebox{\textwidth}{!}{
\begin{tabular}{c|c|cccccccc}
\hline
Up-sampling &  DSC & Aorta& Gallbladder& Kidney(L)& Kidney(R)& Liver& Pancreas& Spleen& Stomach\\
\hline
Bilinear interpolation & 76.15& 81.84 & 66.33&80.12&73.91&93.64&55.04&86.10&72.20\\
Transposed convolution &  77.63 & 84.81&65.96&82.66&74.61&\textbf{94.39}&54.81&89.42&74.41\\
Patch expand & \textbf{79.13} & \textbf{85.47}&\textbf{66.53}&\textbf{83.28}&\textbf{79.61}&94.29&\textbf{56.58}&\textbf{90.66}&\textbf{76.60}\\
\hline
\end{tabular}
}
\centering
\end{table}

\subsubsection{Effect of up-sampling:}
\label{upsampling}

Corresponding to the patch merging layer in the encoder, we specially designed a patch expanding layer in the decoder to perform up-sampling and feature dimension increase. To explore the effective of the proposed patch expanding layer, we conducted the experiments of Swin-Unet with bilinear interpolation, transposed convolution and patch expanding layer on Synapse dataset. The experimental results in the Table~\ref{up} indicate that the proposed Swin-Unet combined with the patch expanding layer can obtain the better segmentation accuracy.

\begin{table}[t!]
\caption{Ablation study on the impact of the number of skip connection}\label{skip}
\footnotesize
\resizebox{\textwidth}{!}{
\begin{tabular}{c|c|cccccccc}
\hline
Skip connection &  DSC & Aorta& Gallbladder& Kidney(L)& Kidney(R)& Liver& Pancreas& Spleen& Stomach\\
\hline
0 &  72.46 & 78.71&53.24&77.46&75.90&92.60&46.07&84.57&71.13\\
1 &  76.43 & 82.53&60.44&81.36&79.27&93.64&53.36&85.95&74.90\\
2 &  78.93 & \textbf{85.82}&66.27&\textbf{84.70}&\textbf{80.32}&93.94&55.32&88.35&\textbf{76.71}\\
3 & \textbf{79.13} & 85.47&\textbf{66.53}&83.28&79.61&\textbf{94.29}&\textbf{56.58}&\textbf{90.66}&76.60\\
\hline
\end{tabular}
}
\centering
\end{table}

\subsubsection{Effect of the number of skip connections:}
\label{skip_num}

The skip connections of our Swin-UNet are added in places of the $1/4$, $1/8$, and $1/16$ resolution scales. By changing the number of skip connections to 0, 1, 2 and 3 respectively, we explored the influence of different skip connections on the segmentation performance of the proposed model. In Table~\ref{skip}, we can see that the segmentation performance of the model increases with the increase of the number of skip connections. Therefore, in order to make the model more robust, the number of skip connections is set as 3 in this work.

\begin{table}[t!]
\caption{Ablation study on the impact of the input size}\label{input}
\footnotesize
\resizebox{\textwidth}{!}{
\begin{tabular}{c|c|cccccccc}
\hline
Input size &  DSC& Aorta& Gallbladder& Kidney(L)& Kidney(R)& Liver& Pancreas& Spleen& Stomach\\
\hline
224 &  79.13& 85.47&66.53&83.28&79.61&94.29&56.58&\textbf{90.66}&\textbf{76.60}\\
384 &  \textbf{81.12} & \textbf{87.07}&\textbf{70.53}&\textbf{84.64}&\textbf{82.87}&\textbf{94.72}&\textbf{63.73}&90.14&75.29\\
\hline
\end{tabular}
}
\centering
\end{table}

\subsubsection{Effect of input size:}
The testing results of the proposed Swin-Unet with $224\times224$, $384\times384$ input resolutions as input are presented in Table.~\ref{input}. As the input size increases from $224\times224$ to $384\times384$ and the patch size remains the same as $4$, the input token sequence of Transformer will become larger, thus leading to improve the segmentation performance of the model. However, although the segmentation accuracy of the model has been slightly improved, the computational load of the whole network has also increased significantly. In order to ensure the running efficiency of the algorithm, the experiments in this paper are based on $224\times224$ resolution scale as the input.

\begin{table}[t!]
\caption{Ablation study on the impact of the model scale}\label{model}
\footnotesize
\resizebox{\textwidth}{!}{
\begin{tabular}{c|c|cccccccc}
\hline
Model scale &  DSC  & Aorta& Gallbladder& Kidney(L)& Kidney(R)& Liver& Pancreas& Spleen& Stomach\\
\hline
tiny &  79.13 & 85.47&66.53&83.28&79.61&\textbf{94.29}&56.58&\textbf{90.66}&\textbf{76.60}\\
base &  \textbf{79.25} & \textbf{87.16}&\textbf{69.19}&\textbf{84.61}&\textbf{81.99}&93.86&\textbf{58.10}&88.44&70.65\\
\hline
\end{tabular}
}
\centering
\end{table}

\subsubsection{Effect of model scale:}
Similar to~\cite{swin}, we discuss the effect of network deepening on model performance. It can be seen from Table.~\ref{model} that the increase of model scale hardly improves the performance of the model, but increases the computational cost of the whole network. Considering the accuracy-speed trade off, we adopt the Tiny-based model to perform medical image segmentation.

\subsection{Discussion}

As we all known, the performance of Transformer-based model is severely affected by model pre-training. In this work, we directly use the training weight of Swin transformer~\cite{swin} on ImageNet to initialize the network encoder and decoder, which may be a suboptimal scheme. This initialization approach is a simple one, and in the future we will explore the ways to pre-train Transformer end-to-end for medical image segmentation. Moreover, since the input images in this paper are 2D, while most of the medical image data are 3D, we will explore the application of Swin-Unet in 3D medical image segmentation in the following research.

\section{Conclusion}

In this paper, we introduced a novel pure transformer-based U-shaped encoder-decoder for medical image segmentation. In order to leverage the power of Transformer, we take Swin Transformer block as the basic unit for feature representation and long-range semantic information interactive learning. Extensive experiments on multi-organ and cardiac segmentation tasks demonstrate that the proposed Swin-Unet has excellent performance and generalization ability. 

%
%
%
%

\bibliographystyle{IEEEtran}
\bibliography{SwinUnet}






\end{document}